\documentclass[aps,prl,twocolumn,amsfonts,amssymb,amsmath,floatfix]{revtex4}

\usepackage{graphicx}
\usepackage{color}
\usepackage{subfigure}
\usepackage{bm}
\usepackage{epstopdf}
\usepackage{comment}

\begin{document}
\title{Interaction-Tuned Dynamical Transitions in a Rashba Spin-Orbit Coupled Fermi Gas}

\author{Juraj Radi\'{c}}
\affiliation{Joint Quantum Institute and Department of Physics, University of Maryland, College Park, Maryland 20742-4111, USA}
\affiliation{Condensed Matter Theory Center, Department of Physics, University of Maryland, College Park, Maryland 20742-4111, USA}

\author{Stefan S. Natu}
\affiliation{Joint Quantum Institute and Department of Physics, University of Maryland, College Park, Maryland 20742-4111, USA}
\affiliation{Condensed Matter Theory Center, Department of Physics, University of Maryland, College Park, Maryland 20742-4111, USA}

\author{Victor Galitski}
\affiliation{Joint Quantum Institute and Department of Physics, University of Maryland, College Park, Maryland 20742-4111, USA}
\affiliation{Condensed Matter Theory Center, Department of Physics, University of Maryland, College Park, Maryland 20742-4111, USA}

\date{\today}

\begin{abstract} 
We consider the time evolution of the magnetization in a Rashba spin-orbit-coupled Fermi gas, starting from a fully-polarized initial state. We model the dynamics using a Boltzmann equation, which we solve in the Hartree-Fock approximation. The resulting non-linear system of equations gives rise to three distinct dynamical regimes with qualitatively different asymptotic behaviors of the magnetization at long times. The distinct regimes and the transitions between them are controlled by the interaction strength: for weakly interacting fermions, the magnetization decays to zero. For intermediate interactions, it displays undamped oscillations about zero and for strong interactions, a partially magnetized state is dynamically stabilized. The dynamics we find is a spin analog of interaction induced self-trapping in double-well Bose Einstein condensates. The predicted phenomena can be realized in trapped Fermi gases with synthetic spin-orbit interactions.  
\end{abstract} 

\maketitle

The physics of spin-orbit coupling is responsible for a wide range of physical phenomena such as atomic spectra, the spin Hall effect and topological insulators \cite{Hasan2010,Spintronics2001, DasSarma2004}.  More recently, the creation of artificial gauge fields ~\cite{Spielman2009,Bloch2013} and spin-orbit coupling~\cite{Spielman2011} in ultra-cold Bose and Fermi systems \cite{Zwierlein2012, Wang2012, Spielman2013} has brought spin-orbit-coupled systems  to the forefront of research in atomic, molecular, and optical physics. Furthermore, cold atoms bring to bear new tools for studying correlated systems. One such tool is the ability to \textit{dynamically} tune the single-particle and many body energy scales \cite{Tiesinga2010}. By varying system parameters adiabatically or diabatically, experimentalists can probe interacting states both in and out of equilibrium. Such  studies have not only provided access to thermodynamic quantities, but have also yielded insights into how correlations develop and spread across a system following a parameter quench, how long-range order is established, and the mechanisms underlying thermalization in isolated interacting systems \cite{Bloch2008, Polkovnikov2011, Bloch2012, Schneider2012, Chin2013}. Here we ask: what is the interplay between spin-orbit coupling and interactions in a gas which is driven out-of-equilibrium? 

%
\begin{figure}
\centerline{
\mbox{\includegraphics[width=1\columnwidth]{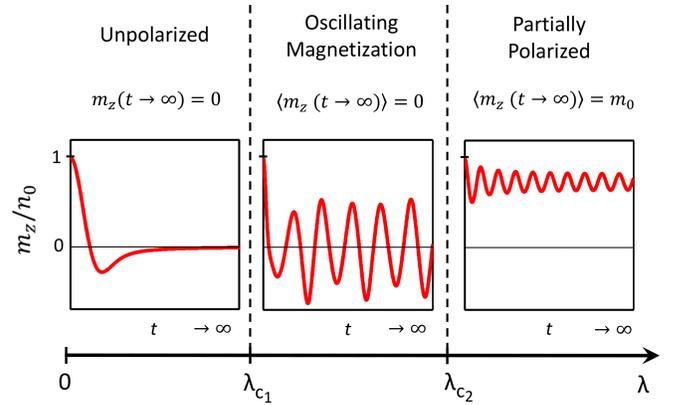}}
}
\caption{(color online)
Schematic plot showing non-equilibrium steady states of an initially spin-polarized Fermi gas in the presence of SOC. 
Here $\lambda=g n_0/\alpha p_{\rm th}$, where $gn_{0}$ parametrizes the interactions (Eq.~\ref{interactions}) and $\alpha p_{\rm th}$ parametrizes the SOC (Eq.~\ref{the_model}). Different types of steady states are separated by transition points $\lambda_{c1}$ 
and $\lambda_{c2}$. For small interactions ($\lambda<\lambda_{c1}$), magnetization decays to zero. For $\lambda_{c1}<\lambda<\lambda_{c2}$ magnetization oscillates forever around zero,  while for large interactions ($\lambda>\lambda_{c2}$) the system becomes partially polarized. 
} 
\label{pdschematic}
\end{figure} 

We study spin dynamics occurring in a weakly interacting, uniform spin-orbit coupled Fermi gas, which is initially spin polarized. We assume that the momentum distribution in the initial state is the classical Maxwell distribution. Interplay between classical motion in the thermal gas, quantum spin degrees of freedom subject to spin-orbit coupling (SOC) and interactions gives rise to interesting dynamical regimes, which are the focus of this work.  Our main results are summarized in Fig.~1: we find that there are three possible distinct steady states which can be labelled as: unpolarized state, oscillating magnetization state, and partially polarized state. Different steady states correspond to different values of the ratio of the interaction strength to the SOC strength (denoted as  $\lambda$), and we find clear transitions between different regimes as we change this $\lambda$.

For weak interactions, the system can be treated within the Hartree-Fock (``collisionless'') approximation (see \cite{KadanoffBaym}), which as shown below, leads to the Boltzmann equation:
\begin{equation}
\frac{\partial \hat{f}}{\partial t} = \frac{i}{\hbar} \left[ \hat{f},\alpha \left( p_x \hat{\sigma}_x + p_y \hat{\sigma}_y \right) 
- \frac{1}{2} g {\bm M} \cdot \hat{\bm \sigma} \right], 
\label{Boltzmann_h}
\end{equation}
where $\hat{f}(\bm{p},t)$ is the Wigner distribution,
$\hat{\sigma}_i$ are Pauli matrices, $\bm{M}=\lbrace m_x, m_y, m_z \rbrace$ is the density-dependent magnetization, $\bm{p}$ is momentum, 
and $\alpha$ and $g$ denote the SOC and interaction strength (details and the derivation 
are discussed later in the text) respectively. 
This equation describes spin-precession for a particle with momentum $\bm{p}$ in
an effective magnetic field, 
$\bm{B}_{\rm eff}=(\alpha p_x -g m_x/2, \alpha p_y -g m_y/2, -g m_z/2)$, 
which involves a combination of SOC and interactions. The fully polarized initial state points along the $z-$ (longitudinal) direction.

If $g=0$, $\bm{B}_{\rm eff}$ points in the x-y (transverse) plane; spins of particles with different momenta precess at different 
frequencies causing dephasing, and the total magnetization decays to zero at long times.
For very large $g$, for a fully polarized initial state, the effective field points close 
to the $z$-axis (Fig. 2). As a result the total magnetization does not decay to zero, but approaches a finite value at long times.
When SOC and interactions are of comparable strength, 
the dynamics leads to a steady state where the magnetization 
shows undamped oscillations about a zero mean-value. This unusual state
is a result of the strongly non-linear character of Eq.~(\ref{Boltzmann_h}). 

In a solid-state material, spin dynamics is primarily influenced by spin-orbit coupling and scattering from impurities and 
phonons \cite{Dyakonov1971, Eliot1954, Yafet1983}. Collisions lead to 
random changes in the electron's momentum, which in turn causes the spin to precess in random directions, leading to Dyakonov-Perel spin relaxation \cite{Dyakonov1971} and 
diffusive dynamics of magnetization \cite{Galitski2007}.

In an ultra-cold gas, the situation is different: there is no disorder, and 
the inter-particle interactions are typically important. In most of the experiments so far, these interactions take the form of a density-density, s-wave contact interaction between two different hyperfine states of a fermionic atom (typically $^{6}$Li or $^{40}$K):
\begin{equation}
\check{V}_{\rm int}=g \int d \bm{r} \check{n}_{\uparrow}(\bm{r}) \check{n}_{\downarrow}(\bm{r}),
\label{interactions}
\end{equation}
where the two hyperfine states are denoted using pseudo-spin variables $\{\uparrow, \downarrow\}$, the interaction parameter $g$ is proportional to the s-wave scattering length $a$ (it can be tuned using Feshbach resonances~\cite{Tiesinga2010}), and $\check{n}_{\uparrow (\downarrow)} (\bm{r})$ is the spin-density operator. S-wave interactions between like fermions are forbidden by the Pauli exclusion principle. 

Depending on the strength of the inter-particle interactions, the gas can be described as either  \textit{collisionless} (Knudsen) or \textit{collision dominated}. The former limit, which is the focus of this work, is achieved when the interactions are so weak, that the timescale for mean-field spin exchange, given by $\tau_{\text{mf}} \sim m/\hbar an_{0}$, is much shorter than the typical collision time $\tau_{\text{coll}} \sim 1/(n_{0} \bar{v} \sigma)$. Here $n_{0}$ is the density, $m$ is the mass, $\bar{v}$ is the average velocity of the particles, and $\sigma = 4\pi a^{2}$ is the scattering cross-section. For typical experimental densities, and temperatures $T \sim T_{F}$, the collisionless regime corresponds to scattering lengths $a \sim 10 a_{\text{B}}$, which yields a $\tau_{\text{coll}} \sim 1$s and a mean-field time $\tau_{\text{mf}} \sim 50$ms. Hence the collisionless limit can be readily explored in experiments \cite{Thomas2008}. 


In this paper we consider a spin-$1/2$ Fermi gas with Rashba spin-orbit coupling:
\begin{equation}
 \hat{H}_0=\frac{{\mathbf p}^2}{2m} + U(\bm{r}) + \alpha \left( \hat{\sigma}_x p_x + \hat{\sigma}_y p_y \right),
\label{the_model}
\end{equation}
where ${\mathbf p}= -i \hbar \bm{\nabla}$ and $U(\bm{r})$ is the external potential. 
This single-particle Hamiltonian can lead to novel spin textures and spin dynamics, even in the absence of interactions \cite{Sau2011, Natu2013, StanescuPRL2007,Wang2010, Ho2011}. 

To study the dynamics of the system, we derive the collisionless Boltzmann equation which describes the evolution of the Wigner distribution:
\begin{equation}
\small{
f_{ij}(\bm{r},\bm{p},t)=\int d\bm{R} e^{i \bm{p} \cdot \bm{R}/\hbar} \Big \langle \check{\psi}^{\dagger}_{i} 
\left(\bm{r}-\frac{\bm{R}}{2},t \right) \check{\psi}_{j} \left(\bm{r}+\frac{\bm{R}}{2},t \right) \Big \rangle,  
}
\label{Wigner}
\end{equation}
which contains all the information about single-particle observables $\left( i,j \in (\uparrow, \downarrow) \right)$. 
For example, total density is 
$n(\bm{r})=n_{\uparrow}(\bm{r})+n_{\downarrow}(\bm{r})=\int d\bm{p}/(2 \pi \hbar)^d 
\left( f_{\uparrow \uparrow} + f_{\downarrow \downarrow} \right)$, 
where $d$ is dimensionality
of the system. 

For spin-orbit coupled fermions with contact interactions, the collisionless Boltzmann equation reads \cite{KadanoffBaym,Halperin2003,Natu2013}:
\begin{equation}
\small{
\begin{split}
&   \frac{\partial \hat{f}}{\partial t}+\frac{\bm{p}}{m} \cdot \nabla_{\bm{r}} \hat{f} 
-\frac{i}{\hbar} \left[ \hat{f},\alpha \left( p_x \hat{\sigma}_x + p_y \hat{\sigma}_y \right) + \hat{V} \right] \\
& \qquad -\frac{1}{2} \lbrace \nabla_{\bm{p}} \hat{f}, \nabla_{\bm{r}} \hat{V} \rbrace  +\frac{1}{2} \alpha \lbrace \partial_x 
\hat{f}, \hat{\sigma}_x \rbrace
+\frac{1}{2} \alpha \lbrace \partial_y \hat{f}, \hat{\sigma}_y \rbrace
=0,
\end{split} 
}
\label{Boltzmann}
\end{equation}
where:
\begin{equation}
\hat{f}(\bm{r},\bm{p},t)= \begin{pmatrix} f_{\uparrow \uparrow}(\bm{r},\bm{p},t) & f_{\downarrow \uparrow}(\bm{r},\bm{p},t) \\ 
f_{\uparrow \downarrow}(\bm{r},\bm{p},t) & f_{\downarrow \downarrow}(\bm{r},\bm{p},t) \end{pmatrix},
\end{equation}
and $\hat{V}$ contains effects of an external potential $U(\bm{r},t)$ and interactions:
\begin{equation}\label{intpot}
\hat{V}(\bm{r},t)=\begin{pmatrix}
U(\bm{r},t)+g n_{\downarrow}(\bm{r},t) & -g n_{-}(\bm{r},t) \\
-g n_{+}(\bm{r},t) & U(\bm{r},t)+g n_{\uparrow}(\bm{r},t)
\end{pmatrix},
\end{equation}
where $n_{\uparrow}=\int d\bm{p}/(2 \pi \hbar)^d f_{\uparrow \uparrow}$, 
$n_{\downarrow}=\int d\bm{p}/(2 \pi \hbar)^d f_{\downarrow \downarrow}$,
$n_{+}=\int d\bm{p}/(2 \pi \hbar)^d f_{\uparrow \downarrow}$ and
$n_{-}=n_{+}^{*}$. The diagonal terms in Eq.~\ref{intpot} represent the direct (Hartree) contribution to the interactions, while the off-diagonal terms represent spin exchange. We define a magnetization vector ${\bm M} = \{m_{x}, m_{y}, m_{z} \} = \{n_{+} + n_{-}, -i \left( n_{+} - n_{-} \right), n_{\uparrow} - n_{\downarrow}\}$, which can be readily probed in ultra-cold atom experiments via spin sensitive phase contrast imaging \cite{Vengalattore2007}. We refer to $m_{z}$ and $\{m_{x}, m_{y}\}$ as the longitudinal and transverse magnetization respectively. 

\begin{figure}
\centerline{
\mbox{\includegraphics[width=\columnwidth]{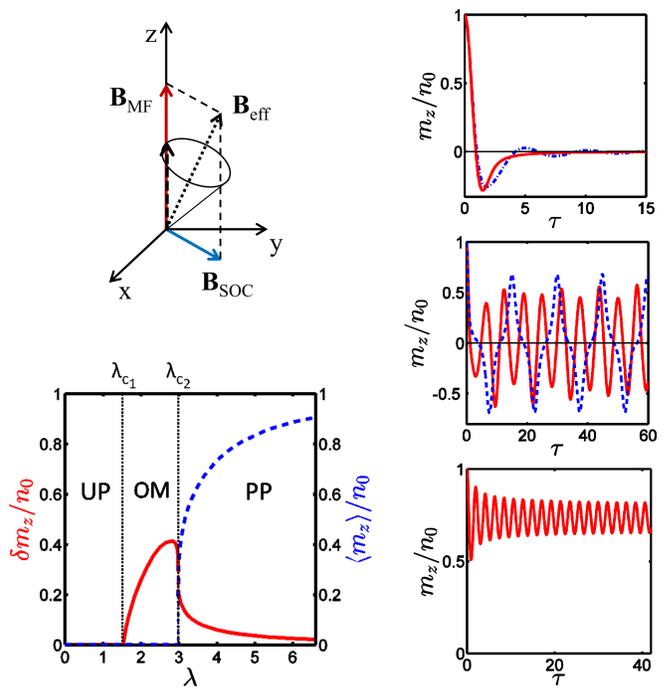}}
}
\caption{\label{magfieldpic} Top-Left: SOC and interactions can be viewed as magnetic fields in the longitudinal (red solid arrow) and transverse direction (blue solid arrow) respectively. The spin precesses around $\bm{B}_{\rm eff}=\bm{B}_{\rm SOC} + \bm{B}_{\rm MF}$. Right Panels: Snapshots of the magnetization for various interaction strengths showing the three different final states: unpolarized, oscillating magnetization and partially polarized state. (Top) Solid red curve is the non-interacting result (see text for analytic formula), and the blue dashed-dotted blue curve has $\lambda = 1.5$. (Center) $\lambda =  2.3$ and the strongly anharmonic curve is $\lambda = 2.977$. (Bottom) $\lambda = 4$. Bottom-Left: The long time-averaged magnetization (dashed blue) and the amplitude of fluctuations about the average $\delta m_{z} = \sqrt{ \langle (m_{z} - \langle m_{z} \rangle)^{2}\rangle}$ (solid red), for different values of $\lambda$. At a first critical value, $\lambda_{c1} = 1.53$, the magnetization develops undamped oscillations about zero, and at a second critical value $\lambda_{c2} = 2.98$, the net magnetization jumps to a non-zero at long times (vertical blue dotted line). 
}
\end{figure}

	In this paper we consider a uniform system ($U(\bm{r}, t) = 0$), where all the relevant quantities become independent of space
(the spatial derivative terms in Eq.~(\ref{Boltzmann}) vanish) and the Boltzmann equation reduces to (\ref{Boltzmann_h}),
where we have conveniently expressed $\hat{V}(t)$ as: $\hat{V}(t)=-g \bm{M}(t) \cdot \hat{\bm{\sigma}}/2$. The effects of a spatially varying potential $U(\bm{r})$ will be discussed later. 
	Spin-orbit coupling can be viewed as a \textit{momentum-dependent} transverse magnetic field $\bm{B}_{\text{SOC}}$, as shown in Fig.~\ref{magfieldpic}. Similarly, the interactions $\hat V$, can be viewed as a \textit{time-dependent} magnetic field $\bm{B}_{\rm MF}(t) = -g\bm{M}(t)/2$ where the time dependence of $\bm{M}$ arises from the density. 
Introducing dimensionless units of momentum $\bm{p}^{\prime}=\bm{p}/p_{\rm th}$ 
($p_{\text{th}}  = \sqrt{2mk_{B}T}$) and time $\tau=t/t_{\rm soc}$ ($t_{\rm soc}=\hbar/\alpha p_{\rm th}$) 
in Eq.~(\ref{Boltzmann_h}), the only parameter in the equation is the ratio of the interaction and SOC strength 
$\lambda=g n_0/ \alpha p_{\rm th}$.

  Our initial state is a non-degenerate, fully polarized gas of spin-$\uparrow$ fermions in thermal equilibrium, at a temperature $T$. We require the temperature to be small enough so that the system can be regarded as a two-level system. In the experiment of Wang \textit{et al.} \cite{Wang2012}, the hyperfine splitting between the states is of order $10~T_{F}$, so this requirement is readily met. The Wigner function reads $f_{\uparrow \uparrow}(\bm{p},t=0)
=\mathcal{N} \exp \left[-\beta \bm{p}^2/(2m) \right]$, $f_{\uparrow \downarrow}(t=0)=f_{\downarrow \uparrow}(t=0)=f_{\downarrow \downarrow}(t=0)=0$, where $\mathcal{N}$ is an overall normalization factor proportional to the total number of particles and
$\beta=1/k_{\rm B} T$. 

As identical spins do not interact, this initial state is stationary in the absence of spin-orbit coupling. We then suddenly switch on the spin-orbit coupling, and numerically integrate Eq.~(\ref{Boltzmann_h}) on a two-dimensional $200 \times 200$ grid in momentum space using the fourth-order Runge-Kutta method. The interaction matrix $\hat{V}$ is sequentially updated at each timestep, and the step size is small enough so that particle number is conserved to very high accuracy. 
All our calculations are done for a two-dimensional system, however the results readily generalize to $3$D, as the Maxwell-Boltzmann distribution is separable in phase space.

In the absence of interactions, the Boltzmann equation (\ref{Boltzmann_h}) reduces to a linear equation, which can be solved \textit{analytically}. The longitudinal magnetization is then $m_{z}(\tau)= n_{0}[1-\sqrt{\pi}\tau e^{-\tau^{2}}\text{Erfi}(\tau)]$, where $\text{Erfi}$ is the imaginary error function.
At short times $\tau \ll 1$, $m_{z} \sim n_{0}(1 - 2\tau^{2} + {\cal{O}}(\tau^{4}))$, while at long times, $m_{z}$ vanishes to zero as $m_z(\tau \rightarrow \infty) \rightarrow -n_0/(2 \tau^2)$. 


The Rashba term in Eq.~(\ref{Boltzmann_h}) couples the spin-$\uparrow$ and $\downarrow$ states by acting as a transverse magnetic field. An atom initially aligned along $\uparrow$ will therefore perform Rabi oscillations about the Bloch sphere. Atoms with different momenta Rabi oscillate at different frequencies, and over time, the total longitudinal magnetization decays to zero. 

The rapid decay of the total magnetization is a property of the thermal gas. At $T=0$, the dominant momenta participating in the dynamics are centered around the Fermi momentum $p_{\rm F}$, which produces magnetization oscillations, in addition to a decay. 
In the $2$D case: $m_{z}(\tau') = n_{0}\Big[\sin(2\tau')/\tau' - \sin^{2}(\tau')/\tau'^{2}\Big]$, where $\tau' = t\hbar/\alpha p_{\rm F}$. The slower decay of the magnetization ($\sim 1/\tau'$) at zero temperature, compared to the non-degenerate gas is due to the larger spread of momenta participating in the dynamics in the thermal case. 

The amount by which the magnetization decays depends on the choice of initial state. For example, for a initial Wigner distribution of the form $\hat{f}(\bm{p},t=0) = \mathcal{N}/2 \exp \left[-\beta \bm{p}^2/(2m) \right]\hat\sigma_{x}$, the average magnetization $m_{x}$ decays to \textit{half} its original value, rather than to zero. 

The situation becomes much more interesting in the presence of interactions, as the Boltzmann equation now becomes a non-linear equation, which has to be solved self-consistently. In Fig.~\ref{magfieldpic}, we plot the effect of interactions on the evolution of the total magnetization. To understand this dynamics, first note that since $n_{+} = n_{-} = 0$ in the initial state, the dynamics does not generate any transverse components of $\bm{M}$ at subsequent times. Exchange interactions therefore do not play an important role in the dynamics described here. The interaction term can  simply be expressed as a time-dependent Zeeman field: $\hat V = \bm{B}_{\text{MF}}(t) \cdot \hat{\bm{\sigma}} = -gm_{z}(t)~\hat{\sigma}_{z}/2$. 

For weak interactions, the dynamics is analogous to the non-interacting case. Spin-orbit coupling leads to spin dephasing on a timescale $t_{\text{soc}} \ll t_{\text{mf}}$, and the longitudinal magnetization $m_{z}$ (and $\bm{B}_{\text{MF}}$) decays to zero. As a result the system becomes effectively non-interacting at long times. 

The steady-state magnetization remains zero until the interactions reach a critical value $\lambda_{c1} = 1.53$. For stronger interactions, the magnetization is still zero on average, but displays large undamped oscillations. As the interactions are increased, the oscillations become strongly anharmonic, and the amplitude and period of the oscillations grows.  Beyond a second critical value $\lambda_{c2} = 2.98$, the average magnetization approaches a finite value at long times (indicated by the blue dotted line). In the vicinity of $\lambda_{c2}$, the oscillation period becomes so long that we are not able to resolve whether the observed jump in the magnetization is merely an artifact of a finite time simulation. As $\lambda \rightarrow 0$, the average magnetization approaches unity, and the amplitude of the oscillations about the average value, monotonically decreases to zero. 


The non-trivial spin dynamics we find is analogous to self-trapping in a  double well Bose condensate, which has been well studied experimentally and theoretically \cite{Chapman2013, Oberthaler2005, Shenoy1997}. If the difference between the on-site interaction energy in each well is small compared to the tunnel splitting, atoms initially prepared in one well undergo sinusoidal Rabi oscillations between the two wells.  For stronger interactions there is a transition to a ``self-trapped" configuration, where the atoms prefer to remain in one of the two wells. 

\begin{figure}
\centerline{
\mbox{\includegraphics[width=1\columnwidth]{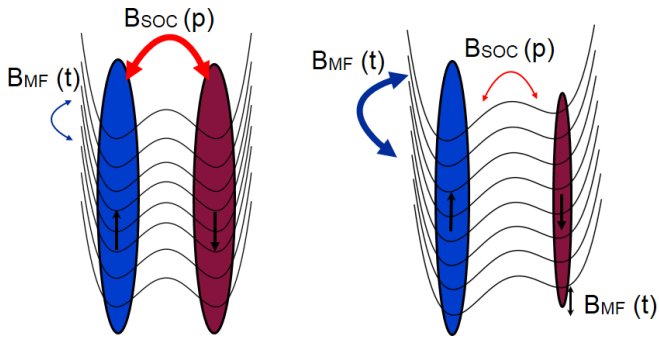}}
}
\caption{\label{doubwell} Our spin system as a collection of double wells, corresponding to spin-$\uparrow$ (blue oval, left well) and spin-$\downarrow$ (red oval, right well) at each momentum. The intra-well tunnelling is provided by the spin-orbit magnetic field $B_{\text{SOC}}(\bm{p})$ (red arrow), which is momentum dependent, while the different wells are coupled to one another via the interaction term $B_{\text{MF}}(t)$ (blue arrow). (Left): Weak interactions (Right) Strong interactions. See the text for details.
} 
\label{pdschematic}
\end{figure}


As shown in Fig.~\ref{doubwell}, our system can be viewed as a collection of ``double wells" (corresponding to the spin-$\uparrow$ and spin-$\downarrow$ states) indexed by their momentum $\bm{p}$. Spin orbit coupling plays the role of the intra-well tunneling, while the role of the on-site interaction energy difference is played by the mean-field term $B_{\text{MF}} = -gm_{z}(t)/2$. Furthermore, as the mean-field term depends on the spin density, it provides the coupling between the double wells, and can thus lead to collective dynamics. For weak interactions ($\lambda \ll 1$), (left panel in Fig.~\ref{doubwell}) the double wells are more or less independent of one another, and spins perform Rabi oscillations between $\uparrow$ and $\downarrow$ states with a frequency proportional to their momentum. This leads to a net cancellation of the total magnetization. For stronger interactions, the inter-well interaction dominates, and spins behave collectively. This leads to coherent, undamped magnetization oscillations. As the interaction strength is increased, the oscillations become anharmonic, as in the double well case \cite{Shenoy1997}. For even stronger interactions (right panel), the initial on-site energy difference between the left and right wells becomes so much larger than the spin-orbit energy, that atoms prefer to remain in one spin state. This results in a non-zero net magnetization on long times, analogous to the self-trapped configuration in the double-well Bose condensate. 

We now discuss the additional physics introduced by harmonic confinement. The trap introduces a new time-scale $t_{\text{trap}} \sim 2\pi/\omega \sim 100$~ms, where $\omega$ is the trapping frequency. The dynamics is different depending on whether $t_{\text{soc}}$ and $t_{\text{MF}}$ are fast or slow compared to the trap period. If $t_{\text{soc}} \sim t_{\text{MF}} \ll t_{\text{trap}}$, we locally recover the physics of the homogeneous case, 
\textit{i.e.} the magnetization dynamics at each point in the trap will be the same as in the uniform case, however the parameter $\lambda$ will vary in space as $\lambda(r)=g n(r)/(\alpha p_{\rm th})$, where $n(\bm{r})$ is the local density. If the timescale for spin-orbit coupling and the trap are comparable, the total magnetization displays collapse and revival dynamics \cite{Natu2013, StanescuPRL2007}. Recent proposals of using pulsed magnetic fields to generate tunable spin-orbit couplings \cite{BAnderson2013, Ueda2013} and the demonstration of nearly uniform, ``box" potentials \cite{Hadzibabic2013}, will enable experimentalists to explore this wide regime of parameters. 

In summary, we have shown that out-of-equilibrium dynamics of quantum spins in an otherwise classical gas leads to
to non-trivial steady states and associated dynamical transitions. The steady states we describe are far from thermal equilibrium, and their stability relies on the absence of collisions. In cold atom experiments, the strength of interactions can be made weak enough such that 
the number of collisions throughout the duration experiment are negligible. However, on long enough times, collisions will eventually lead to thermalization and a dissapearance of the steady states we describe. Our numerical simulations indicate that the steady states and dynamical transitions we find also occur for more general forms of SOC. 


Experiments on spin-orbit coupled Fermi gases are already beginning to explore spin dynamics driven by SOC \cite{Wang2012}.  A systematic investigation of how spin dynamics is influenced by temperature and interactions is an active area of study \cite{Tokatly2013, Yu2013}, and is of considerable relevance to future experiments on strongly correlated spin-orbit coupled quantum gases.

\textit{Acknowledgements.---} This work was supported by  ARO-MURI (J.R. and S.N.), JQI-NSF-PFC (S.N.), AFOSR-MURI (S.N.), and US-ARO (V.G.).

\bibliography{SpinDyn_references}

\end{document}